\documentclass[aps,floatfix]{revtex4}

\usepackage{graphicx}
\usepackage{dcolumn}
\usepackage{bm}
\usepackage{amsmath}
\usepackage{graphicx}
\usepackage{listings}
\usepackage[T1]{fontenc}
\usepackage{xcolor}
\usepackage{lmodern}
\usepackage{listings}
\usepackage[caption=false]{subfig}

\begin{document}

\title{{\bf Thermodynamics of FRW Universe With Chaplygin Gas Models}}

\author{\bf{Samarjit Chakraborty and Sarbari Guha}}

\affiliation{\bf Department of Physics, St.Xavier's College (Autonomous), Kolkata 700016, India}

\maketitle
\section*{Abstract}
In this paper we have examined the validity of the generalized second law of thermodynamics (GSLT) in an expanding Friedmann Robertson Walker (FRW) universe filled with different variants of Chaplygin gases. Assuming that the universe is a closed system bounded by the cosmological horizon, we first present the general prescription for the rate of change of total entropy on the boundary. In the subsequent part we have analyzed the validity of the generalised second law of thermodynamics on the cosmological apparent horizon and the cosmological event horizon for different Chaplygin gas models of the universe. The analysis is supported with the help of suitable graphs to clarify the status of the GSLT on the cosmological horizons. In the case of the cosmological apparent horizon we have found that some of these models always obey the GSLT, whereas the validity of GSLT on the cosmological event horizon of all these models depend on the choice of free parameters in the respective models.

\bigskip

KEYWORDS: Cosmology; Chaplygin gas; Universal thermodynamics.

\section{Introduction}
Thermodynamics plays a very crucial role both in cosmological analyses as well as in the General Theory of Relativity.
The semi-classical description of black hole physics tells us that a black hole emits thermal radiation and behaves like a black body. This led to the successful description of a black hole as a thermodynamic system \cite{SWH1}. The introduction of the Bekenstein-Hawking entropy on the black hole event horizon yielded the complete development of the laws of black hole thermodynamics. Bekenstein had to assign an entropy function to a black hole in order to save the second law of thermodynamics (SLT) from becoming erroneous on the black hole horizon \cite{Bekenstein}. The temperature and the entropy of the black hole are proportional to the surface gravity on the horizon and the area of the horizon, respectively. Hence these parameters are related to the geometry of the black hole horizon. Moreover, the temperature, the entropy and the mass of the black hole were found to satisfy the first law of thermodynamics \cite{BCH}.

All these prompted physicists to search for a possible connection between black hole thermodynamics and the gravitational field equations. Jacobson \cite{Jacobson} was the first to derive the Einstein field equations from the proportionality of the black hole entropy and the horizon area together with the fundamental relation $\delta Q = TdS$, claiming that this relation is valid for all local Rindler causal horizons through each space time point,  with $\delta Q$ and $T$ as the energy flux and Unruh temperature seen by an accelerated observer just inside the horizon. Subsequently, Hayward \cite{Hayward} derived a unified first law of black-hole dynamics and relativistic thermodynamics in spherically symmetric general relativity. It was Padmanabhan \cite{Paddy} who formulated the first law of thermodynamics on ``any'' horizon for a general static spherically symmetric space time, starting from the Einstein equations. Thus the equivalence of the laws of thermodynamics with the analogous laws of black hole mechanics on one side and the Einstein equations of the classical theory of gravity on the other side, revealed a strong connection between quantum physics and gravity.

In the same way, on the cosmological scale, the SLT can be implemented by assuming that the universe is a closed system bounded by some horizon, preferably the cosmological apparent horizon. Applying the first law of thermodynamics to the apparent horizon of a FRW universe and considering the Bekenstein entropy on the apparent horizon, Cai and Kim \cite{caikim} derived the Friedmann equations for a universe with any spatial curvature. They used the entropy formulae for the static spherically symmetric black hole horizons in Gauss-Bonnet gravity and in Lovelock gravity, to obtain the Friedmann equations in these theories. Paranjpe et al \cite{PSP} showed that the field equations for the Lanczos-Lovelock action in a spherically symmetric spacetime can also be expressed in the form of the first law of thermodynamics. Akbar and Cai \cite{AC1} extended the work of Cai and Kim to the cases of scalar–tensor gravity and $f(R)$ gravity, and subsequently showed that \cite{AC2} the Friedmann equation of a FRW universe can be rewritten as the first law of thermodynamics on the apparent horizon of the universe and extended their procedure to the Gauss-Bonnet and Lovelock gravity. Cai and Cao \cite{CC} showed that the unified first law proposed by Hayward for the outer trapping horizon of a dynamical black hole could be applied to the apparent horizon of the FRW universe for the Einstein theory, Lovelock theory, and the scalar-tensor theories of gravity.

Although the cosmological event horizon does not exist in the big bang model of standard cosmology, but in a general accelerating universe dominated by dark energy, the cosmological event horizon separates out from the apparent horizon. Considering the physically relevant part of the Universe to be bounded by the dynamical apparent horizon, Wang et al \cite{WGA} showed that although both the first and the second laws
of thermodynamics are satisfied in such a case, but if the boundary of the Universe is assumed to be the cosmological event horizon, then both these laws break down at the event horizon, if the usual definition of temperature and entropy as applicable to the apparent horizon is extended to the event horizon. According to them, the first law may apply only to variations between nearby states of local thermodynamic equilibrium whereas the event horizon reflects the global properties of spacetime.

The conditions of validity of the generalized second law of gravitational thermodynamics in the phantom-dominated era of the flat FRW universe, was examined by Sadjadi \cite{Sadjadi}. Considering a homogeneous and isotropic universe, filled with perfect fluid having an arbitrary equation of state, Mazumder and Chakraborty \cite{MC1,MC2} have shown the validity of the GSLT of the universe with the event horizon as the boundary assuming the first law of thermodynamics, with some restrictions on the matter. Jamil et al \cite{JSS} investigated the validity of the GSLT in the cosmological scenario where dark energy interacts with both dark matter and radiation. They calculated separately the entropy variation for each fluid component and for the apparent horizon, and showed that the GSLT is always and generally valid, independently of the specific form of the equation of state (EOS) parameters of the fluids and of the background geometry. Tian and Booth \cite{TB} reexamined the thermodynamics of the Universe by requiring its compatibility with the holographic type gravitational equations which govern the dynamics of both the cosmological apparent horizon and the entire Universe. They proposed possible solutions to the existing problems regarding the temperature of apparent horizon and the evolution of cosmic entropy.

Yang et al \cite{YANG} showed that for a constant EOS of dark energy, the allowed interval of the EOS parameters for the validity of the GSLT has to be $ w_{D}\geq-1 $, in a universe enveloped by the apparent horizon and containing a Schwarzschild black hole. Xing et al \cite{XING} showed the validity of the thermodynamical properties of the universe in a new parametric model of dark energy with the equation of state $w = w_{0}+w_{1}.z(1+z)/(1+z^{2})$. In the spatially homogeneous and isotropic universe, assuming that the temperature and entropy in cosmology is as in a black hole, they examined the thermodynamical properties of the universe bounded by the apparent horizon and the event horizon respectively. They found that the first and the second laws of thermodynamics are valid inside the apparent horizon, while they break down inside the event horizon. Rani et al investigated the validity of the GSLT for a model of pilgrim dark energy interacting with cold dark matter in the frame-work of dynamical Chern-Simons modified gravity in a nonflat FRW universe \cite{RANI}. Sharif et al \cite{SZ} analyzed non-equilibrium aspects of thermodynamics on the apparent horizon of FRW universe in $f(R, T)$ gravity along with the validity of GSLT. Cardone et al studied \cite{CAR} two different dark energy models namely the Barboza-Alcaniz parameterization and the phenomenologically-motivated Hobbit model in the context of the GSLT. Iqbal et al \cite{AYE} investigated the validity of GSLT of the Ricci-Gauss-Bonnet dark energy and cold dark matter bounded by the apparent horizon and event horizon in flat FRW universe.
 Other authors also analyzed the GSLT in various theories and dark energy models like in \cite{MOR1,MOR2,LYM,SSS,JAW}.


Izquierdo and Pavon \cite{IP} explored the thermodynamics of dark energy by assuming the existence of the observer's event horizon in accelerated universes. They found that except for the initial stage of Chaplygin gas dominated expansion, the GSLT is valid in all such cases. The validity of the second law in an expanding Godel-type universe filled with generalized Chaplygin gas interacting with cold dark matter has also been examined \cite{godel}. Sharif and Saleem \cite{SS1} studied the validity of the GSLT in the presence of non-interacting magnetic field and new modified Chaplygin gas with FRW universe. Bamba et al. discussed the viability of the Generalized Chaplygin gas (GCG) as an alternative to $ \Lambda $CDM model to explain the origin of both Dark matter and Dark energy in a single fluid equation. They also discussed how the matter perturbation grows and how the sound speed limits the magnitude of free parameter $ \alpha $ in the EOS of GCG \citep{bamba}. Karami et al \cite{KAR} investigated the validity of the GSLT in a non-flat FRW universe in the presence of the interacting generalized Chaplygin gas with the baryonic matter, where the universe is assumed to be enclosed by the dynamical apparent horizon. Bandyopadhyay \cite{TB1} showed the validity of the GSLT in the braneworld scenario with induced gravity and curvature correction terms along with a dark energy component, namely, the Modified Chaplygin Gas on the 3-brane together with a perfect fluid as the dark matter.

It is therefore evident that although the apparent horizon is physically much more relevant to work with in a dynamical situation, but the event horizon has also its own importance. As we know that in a dynamically evolving universe or a black hole, both of these horizons are present, so it is justified to check for the validity of the generalized second law of thermodynamics (GSLT) on these horizons for a universe filled with various types of matter and/or energy, which in our case of study is a fluid like the Chaplygin gas. Chaplygin gas models are very versatile and useful cosmological models suitable for representing the different phases of evolution of the universe. In fact, the necessity of a model which can explain the evolutionary history of the universe successfully, led to the birth of the Chaplygin gas cosmology. Since the Chaplygin gas models can describe the accelerating expansion of the universe in the current epoch, hence they provide us a robust model for the mysterious Dark Energy. It is therefore quite prudent for one to compare the different Chaplygin gas models from a thermodynamic point of view, to identify the suitability of the different models in this group and hence comment on their merits. For this purpose we have examined the validity of the GSLT both on the cosmological apparent horizon and the cosmological event horizon for the different Chaplygin gas models. As each model in this group is distinct from the other, we obatin different cosmological consequences for the validity of the GSLT on both the horizons in these models.

For the analysis of the cosmological apparent horizon, we have considered the Kodama-Hayward temperature because the Kodama-Hayward surface gravity is more relevant for the description of dynamical horizons \cite{Faraoni}. In the case of the Variable modified Chaplygin gas, we have already determined the temperature of the FRW universe \cite{CGP} in another paper. This temperature is the bulk temperature. In this paper, after calculating the Kodama-Hayward temperature of the VMCG dominated FRW universe for the apparent horizon, we have compared these two types of temperatures to see how their behaviour affects the thermodynamics of the universe. To the best of our knowledge a comparison of this kind have not been done earlier. Further, we want to point out that our approach is much more general compared to other works as we did not assume any specific definition of surface gravity (i.e. temperature) for our analysis in the case of the cosmological event horizon. The analysis of generalized thermodynamics of FRW universe for models like the Variable modified Chaplygin gas (VMCG), New Variable modified Chaplygin gas (NVMCG), Generalized cosmic Chaplygin gas (GCCG), and Modified cosmic Chaplygin gas (MCCG) on both the cosmological horizons is also a completely new study. This will help further analysis on such models in future. 

The plan of our work is as follows: in Section II we present a brief review of the various Chaplygin gas models which we have analyzed in this paper. This is followed by a general description of the theory of gravitational thermodynamics in Section III. In Sections IV and V, we analyze the criterion for which the GSLT will be valid on the cosmological apparent horizon and the cosmological event horizon, respectively, in the case of FRW universes filled with various types of Chaplygin gases. We follow up with some useful discussions in the penultimate section and end up with the conclusions in section VII.

\section{The Chaplygin gas models}
Two major problems of modern cosmology are those concerning Dark Energy and Dark Matter. Dark matter is the invisible mass in the universe or some invisible source of gravity which constitutes approximately $23$ percent of the composition of the observable universe. We also know that the universe is accelerating in its current state of expansion, an effect which is attributed to the presence of Dark Energy (constituting approximately $70$ percent of the observable universe). Although there is no clear understanding about the exact nature of this component, there are different models for explaining its effect. This led to the proposal of various types of exotic fluids as the matter content of the universe. An interesting type of such a fluid is the Chaplygin gas. It is found that some of the Chaplygin gas models can successfully describe all three phases of evolution of the universe, which therefore makes them very useful for cosmological studies. Hence it is necessary to examine the status of the GSLT on the cosmological horizons of FRW universes with matter content in the form of different variants of Chaplygin gas. Below we briefly present the Chaplygin gas models which we have analyzed in this work.

The equation of state for the Chaplygin gas \cite{CG1} is given by
\begin{equation}\label{cg_eos}
p=-B/\rho,
\end{equation}
where $ B $ is a positive constant, $ p $ is the pressure of the fluid, and $ \rho $ is the energy density. The Generalized Chaplygin gas (GCG) \cite{GCG}, is represented by the equation of state
\begin{equation}\label{gcg_eos}
p=-B/\rho^{\alpha},
\end{equation}
where $ \alpha $ is a positive constant lying within the range $0\leq \alpha \leq 1$. For the Modified Chaplygin gas (MCG) model \cite{MCG1}, the equation of state (EOS) is
\begin{equation}\label{mcg_eos}
p=A\rho-B/\rho^{\alpha},
\end{equation}
where $ A $ and $B$ are positive constants. The model which is further generalized is the Variable Modified Chaplygin gas (VMCG) \cite{VMCG1} with the EOS
\begin{equation}\label{vmcg_eos}
p=A\rho-B(a)/\rho^{\alpha},
\end{equation}
where $ B(a)=B_{0}a^{-n}=B_{0}V^{-n/3}$ is a function of the cosmological scale factor $ a $ of the FRW universe, $ B_{0} $ is a positive constant, $ n $ is any constant, and we have assumed $ V=a^{3} $ for the FRW universe.

Advancing further, we have the New Variable Modified Chaplygin gas (NVMCG) \cite{CD} with the EOS
\begin{equation}\label{nvmcg_eos}
p=A(a)\rho-B(a)/\rho^{\alpha},
\end{equation}
where $A(a)=A_{0}a^{-m}$, $ B(a)=B_{0}a^{-n} $ are functions of the cosmological scale factor $ a $, with $A_{0}, B_{0}, m $ as positive constants, $ n $ is any constant, and $ 0\leq \alpha \leq 1 $. The expression for energy density is \cite{CD}
\begin{align}
\rho&=a^{-3}e^{\frac{3A_{0}a^{-m}}{m}} \bigg[c_{0}+\frac{B_{0}}{A_{0}}\left( \frac{3A_{0}(1+\alpha)}{m} \right)^{\frac{3(1+\alpha)+m-n}{m}} \times \Gamma \left(\frac{n-3(1+\alpha)}{m},\frac{3A_{0}(1+\alpha)a^{-m}}{m}\right)\bigg]^{\frac{1}{1+\alpha}},
\end{align}
where $ \Gamma(x,y) $ is the upper incomplete gamma function and $ c_{0} $ is the integration constant.

Then comes the Generalized Cosmic Chaplygin gas (GCCG) \cite{GCCG} with the EOS
\begin{equation}\label{gccg}
p=-\rho^{-\alpha}[c+(\rho^{\alpha+1}-c)^{-w}],
\end{equation}
where $ c=\frac{E}{1+w} -1 $, and $ E $ can take both positive or negative constant values under the condition $ -L< w < 0 $, where $ L $ is a positive definite constant which can be larger than unity. The expression for the energy density in this case is
\begin{equation}
\rho=[c+(c_{1}NV^{-N}+1)^{\frac{1}{w+1}}]^{\frac{1}{\alpha+1}},
\end{equation}
where $c_{1}  $ is an arbitrary integration constant, and $ N=(1+\alpha)(1+w) $.

Finally, we have the Modified Cosmic Chaplygin gas (MCCG) with the EOS \cite{MCCG}
\begin{equation}\label{mccg_eos}
P=A\rho-\rho^{-\alpha}[(\rho^{\alpha+1}-C)^{-\gamma} + C],
\end{equation}
where $ 0<\alpha\leq 1 $, $ -b<\gamma<0 $ and $ b\neq 1  $. Here the parameter $ C=\frac{Z}{\gamma +1} -1 $, where $ Z $ is an arbitrary constant, and $ A $ is a positive constant. In the above EOS, if $ A\rightarrow 0 $, then we arrive at the EOS of GCCG. Using thermodynamic identity and binomial approximation, we obtain the approximate form of energy density of the MCCG as \cite{SA}:
\begin{equation}
\rho=\left[ \dfrac{C+(-C)^{-\gamma}+(\frac{\varepsilon}{V})^{M}}{A+1+\gamma(-C)^{-\gamma-1}}\right]^{\frac{1}{1+\alpha}},
\end{equation}
where $ \varepsilon=d(A+1)^{\frac{1}{M}} $, $ M=(1+\alpha)(1+A) $, and $ A+1+\gamma(-C)^{-\gamma-1}\neq 0 $. Here $ d $ is the constant of integration obtained during the calculation of energy density.

These expressions will be used in our subsequent calculations.

\section{Thermodynamic analysis}

We know that the Friedman equations can be written on a dynamical horizon in the form of the first law of thermodynamics \cite{caikim,AC1}
\begin{equation}
-dE_{H}=T_{H}dS_{H},
\end{equation}
where $ dE_{H} $ is the energy flowing across the horizon, $dS_{H}$ is the change of horizon entropy because of it, and $T_H$ is the horizon temperature.

Let us presume that the first law of gravitational thermodynamics holds on the cosmological horizons, and based on that premise the GSLT can be introduced in the form
\begin{equation}
\frac{dS_{T}}{dt}=\frac{dS_{b}}{dt}+\frac{dS_{H}}{dt}>0,
\end{equation}
where $ S_{T} $ is the total entropy, $ S_{H} $ is the horizon entropy and $ S_{b} $ is the bulk fluid entropy. Therefore, ultimately the GSLT is the direct extension of the SLT which says that the total entropy of the universe plus the cosmological horizon entropy should always increase.

\subsection{General formalism}

Let us assume that the universe bounded by the cosmological horizon is filled with a fluid {\color{blue}{of}} energy density $ \rho $ and pressure $ p $. The energy conservation relation is given by

\begin{equation}
\dot{\rho}+3H(p+\rho)=0,
\end{equation}
where $H$ is the Hubble parameter. The Einstein field equations for homogeneous, isotropic, flat FRW universe are
\begin{align}
3H^2 =\rho, \\
2 \dot{H}=-(p+\rho).
\end{align}

Assuming that the first law of thermodynamics holds on the cosmological horizon, we can write
\begin{equation}\label{firstlaw}
-dE_{H}=T_{H}dS_{H},
\end{equation}
where $ T_{H} $ is the temperature of the horizon, $dE_{H}$ is the amount of energy crossing the horizon in time $ dt $, and $dS_{H}$ is the amount of entropy change of the universe due to it. If $\dot{\rho}$ is the corresponding rate of change of the energy density of the universe, then we can write
\begin{align}
dE_{H}&=\frac{4\pi R_{H}^{3} \dot{\rho}dt}{3} \nonumber \\
&=-4\pi R_{H}^{3}H(p+\rho)dt.
\end{align}
Substituting the above expression in the equation (\ref{firstlaw}) depicting the first law of thermodynamics, we obtain the rate of change of horizon entropy as
\begin{equation}
\frac{dS_{H}}{dt} =\frac{4\pi R_{H}^{3} H(p+\rho)}{T_{H}}.
\end{equation}
We now use the Gibbs equation in the bulk to get the entropy of the fluid bounded by the horizon in the form
\begin{equation}
T_{H}dS_{b}=dE_{b} +pdV,
\end{equation}
where our underlying assumption is that the bulk temperature is equal to the horizon temperature ($ T_{H}=T_{b} $), i.e. the bulk and the horizon surface are in thermal equilibrium. Substituting $ V = 4\pi R_{H}^{3}/3  $ and $ E_{b} = 4 \pi R_{H}^{3}\rho/3 $ in the Gibbs relation, we have
\begin{equation}
T_{H}dS_{b} = (4\pi R_{H}^{2}\rho \dot{R}_{H} +\frac{4}{3}\pi R_{H}^{3}\dot{\rho}+4\pi p R_{H}^{2}\dot{R}_{H})dt.
\end{equation}
Therefore the rate of change of the fluid entropy is
\begin{equation}
\frac{dS_{b}}{dt}=\frac{4\pi R_{H}^{2} (p+\rho)(\dot{R_{H}}-H R_{H})}{T_{H}},
\end{equation}
and the rate of change of total entropy is
\begin{align}
\frac{dS_{T}}{dt} =\frac{d(S_{H}+S_{b})}{dt}   =\frac{4\pi R_{H}^{2}(p+\rho)\dot{R}_{H}}{T_{H}}.
\end{align}

\subsection{Cosmological apparent horizon}

It is known that for a spatially flat FRW universe, the apparent horizon and the Hubble horizon coincides \cite{AC2}. The area radius of the cosmological apparent horizon is given by

\begin{equation}\label{R_AH}
R_{AH}(t)=\frac{1}{H(t)}.
\end{equation}
Now for a flat FRW universe, the apparent horizon evolves according to the relation
\begin{align}
\dot{R}_{AH}(t) = -H \dot{H} R_{AH}^3 = \frac{(p+\rho)}{2H^2}.
\end{align}

Using the above relation in the expression for rate of change of total entropy, we obtain
\begin{equation}
\frac{dS_{T}}{dt}= \frac{4\pi (p+\rho)^2}{2 H^{4}T_{AH}},
\end{equation}
which is a positive quantity, and hence our usual expectation is that the GSLT will always be valid on the apparent horizon when the bulk fluid and the horizon are in thermal equilibrium.

\subsection{Cosmological event horizon}

The proper radius of the event horizon in the FRW universe is
\begin{equation}\label{R_EH}
R_{EH}(t)= a(t)\int^{+\infty}_{t} \frac{dt'}{a(t')},
\end{equation}
where $a(t)$ is the scale factor of the expanding universe. We know that if the above integral converges, then the universe will have an event horizon, and the equation according to which the cosmological event horizon evolves is given by
\begin{equation}
\dot{R}_{EH} =HR_{EH}-1 .
\end{equation}
Using the above relation in the expression for rate of change of  total entropy, we get
\begin{align}
\frac{dS_{T}}{dt}&=\frac{4\pi R_{EH}^{2}}{T_{EH}}(p+\rho)(HR_{EH}-1) \nonumber \\
&= \frac{4\pi R_{EH}^{2}H}{T_{EH}}(p+\rho)(R_{EH}-R_{AH}).
\end{align}

\section{Validity of GSLT on the Cosmological apparent horizon of various Chaplygin gas models}

Lets us now examine the status of the GSLT on the cosmological apparent horizon of FRW universes filled with the different variants of the Chaplygin gas listed in section II.

\subsection{VMCG}

\begin{figure}[ht]
 \includegraphics[width=0.42\textwidth]{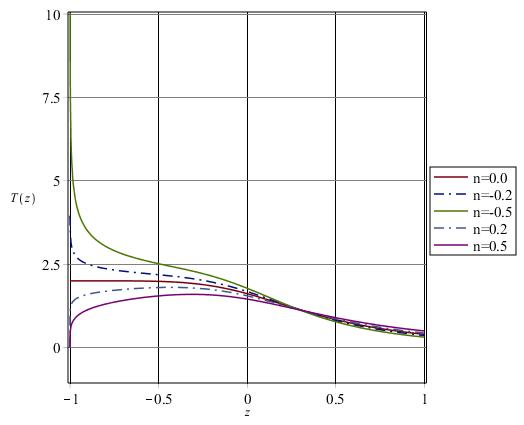}
 \caption{Variation of Kodama-Hayward temperature $T(z)$ on the AH for VMCG dominated FRW universe as a function of $ z $ for different values of $ n $.}
\label{khtemp}
\end{figure}

The Kodama-Hayward temperature (KHT) of the cosmological apparent horizon for a spatially flat FRW universe is given by \cite{BH}
\begin{equation}\label{Kodama}
k_B T = \left( \frac{\hbar G}{c} \right)\frac{(\rho - 3p)}{3H}.
\end{equation}
Here we can rewrite $H$ in terms of the redshift $z$. As (\ref{Kodama}) involves the pressure and energy density of the fluid, it captures the cosmological essence of the model. But this pressure and energy density are related by the EOS of the VMCG given by (\ref{vmcg_eos}) where $B(a)=B_{0}a^{-n}$. Thus we have studied the variation of the KHT on the Apparent Horizon (AH) for VMCG dominated FRW universe as a function of $z$ for different values of $n$. This is shown in FIG.\ref{khtemp}

\begin{figure}[ht]
\includegraphics[width=0.42\textwidth]{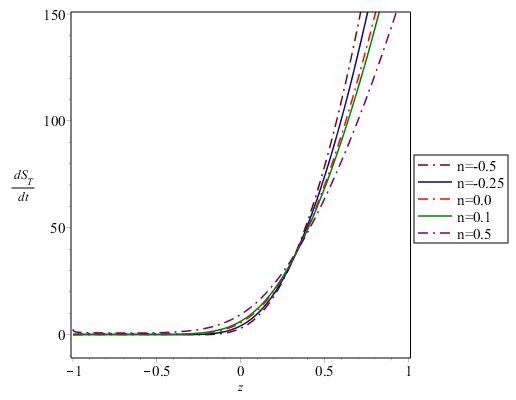}
\caption{Plot showing the variation of $ \dfrac{dS_{T}}{dt} $ with respect to the free parameter $n$ on the cosmological apparent horizon (using the Kodama temperature of the AH) for VMCG dominated FRW universe, validating the GSLT on the AH.}
\label{label-a}
\end{figure}

Using the expression of KHT on the cosmological apparent horizon, we have also studied the variation of $ \dfrac{dS_{T}}{dt} $ with respect to the free parameter $n$ for the VMCG dominated FRW universe, which is shown in FIG.\ref{label-a}. From this figure, it is evident that the total entropy on the AH always increases for different values of the parameter $ n $. Thus the GSLT is always valid on the apparent horizon of the VMCG dominated FRW universe when we consider the Kodama temperature of the horizon.

In a separate paper \cite{CGP} we have already determined the expression for the bulk temperature of the VMCG in a FRW universe.
It is therefore interesting to compare the validity of GSLT on the AH of the VMCG dominated FRW universe determined in terms of the KHT of the AH with the validity determined in terms of the temperature of the FRW universe filled with VMCG matter. For this purpose we have used the expression of the temperature of the VMCG dominated FRW universe which we have calculated in {\textbf{the}} paper \cite{CGP}. Substituting the EOS of the VMCG (given in equation \ref{vmcg_eos}) in the thermodynamic identity $\left(\frac{\partial U}{\partial V}\right)_s = -P $, we obtain the relation
\begin{equation}\label{07}
\left(\frac{\partial U}{\partial V}\right)_s = -A(U/V)+ B_0 V^{-n/3} (V/U)^{\alpha}.
\end{equation}
From equation (\ref{07}) the energy density is determined accurately up to an integration constant
\begin{equation}\label{08}
 \rho = \frac{1}{a^\frac{n}{1+\alpha}}\left[(1+\alpha)B_0/N + C / a^{3N}\right]^\frac{1}{1+\alpha},
\end{equation}
where $N=(A+1)(1+\alpha)-n/3$, and $C$ is the integration constant. The expression of energy density for the VMCG dominated FRW universe, as a function of scale factor, is obtained as
\begin{equation}\label{09}
\rho(a)=\dfrac{\rho_{0}}{a^{\frac{n}{1+\alpha}}}\left[\Omega_x + (a_0^{n}-\Omega_{x})(a_0/a)^{3N}\right]^\frac{1}{1+\alpha},
\end{equation}
where we defined the dimensionless parameter
\begin{equation}\label{10}
\Omega_x =\dfrac{(1+\alpha)B_0}{N \rho_{0}^{1+\alpha}}.
\end{equation}
Finally the temperature $ T(z) $ of this VMCG universe as a function of the redshift $z$ is obtained in the form \cite{CGP}
\begin{align} \label{T_z1}
\left. T(z)\right. =\frac{T_0(z+1)^{3N(1+\frac{\alpha}{1+\alpha}) +3A}(\frac{1}{\Omega_x})^{\frac{\alpha}{1+\alpha}}}{[1+(z+1)^{3N}(\frac{a_0^n}{\Omega_{x}}-1)]^\frac{\alpha}{1+\alpha}} \times\frac{[\frac{1}{\Omega_x}(1-\frac{n}{3N})-\frac{1}{a_0^n}]a_0^{n(1+\frac{\alpha}{1+\alpha})}}
 {[(1-\frac{n}{3N})(\frac{a_0^n}{\Omega_{x}}-1)(z+1)^{3N}-\frac{n}{3N}]}.
\end{align}

\begin{figure}[ht]
 \includegraphics[width=0.45\textwidth]{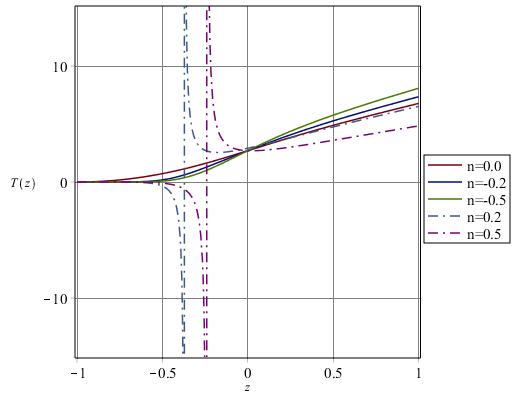}
 \caption{Variation of $T(z)$ as a function of $ z $ for different values of $ n $ for the VMCG dominated FRW universe.}
\label{vmcgtempfuture}
\end{figure}

The temperature of the VMCG dominated FRW universe is depicted in FIG.\ref{vmcgtempfuture} as a function of redshift. It can be seen that in the future epoch, the measure of temperature becomes negative and has an infinite discontinuity, which clearly indicates that the VMCG model is thermodynamically unstable for positive values of the parameter $ n $. This is consistent with the conclusions derived by Panigrahi and Chatterjee in \cite{vmcg}.

\begin{figure}[ht]
    \centering
    \subfloat[Subfigure 1 list of figures text][]
        {
        \includegraphics[width=0.45\textwidth]{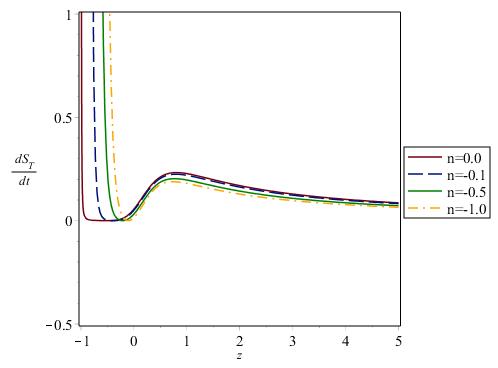}
        \label{fig:subfig1}
        }
    \subfloat[Subfigure 2 list of figures text][]
        {
        \includegraphics[width=0.43\textwidth]{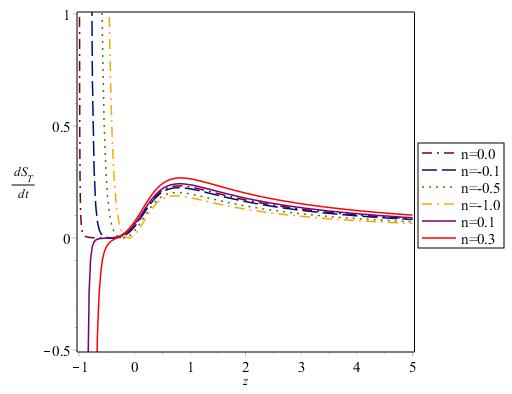}
        \label{fig:subfig2}
        }
    \caption{Variation of $ \dfrac{dS_{T}}{dt} $ wrt the free parameter $n$ preserving the validity of GSLT on the cosmological apparent horizon (using temperature of FRW universe dominated by VMCG).}
    \label{fig2}
\end{figure}

In the next two figures (shown in FIG.\ref{fig2}), we have used the temperature of VMCG dominated FRW universe in place of the horizon temperature for the near equilibrium scenario. In this case we observe that there is a clear difference from FIG.\ref{label-a}. For negative $ n $, the GSLT is always valid on the AH but for positive $ n $, the GSLT gets violated in the future epoch (i.e. for negative redshift).

As we have used the VMCG temperature in our entropy calculation, it is natural for us to come across a scenario in which the total entropy decreases for the range of positive values of $ n $. In actuality, this reflects the inherent thermodynamic behaviour of the model itself in the cosmological context. In FIG.\ref{fig2}(b), the negative rate of change of total entropy for $ n>0 $ is the result of the thermodynamic instability of VMCG for $ n>0 $ \cite{vmcg}. This is why we get to see different nature of the plots in FIG.\ref{label-a} and FIG.\ref{fig2}, respectively, when we use the two different temperatures (i.e. KHT and the temperature of VMCG, respectively).

In this context it may be noted that Chen et al. \cite{CLX} have applied cosmological constraints on the VMCG model using the Markov chain Monte Carlo (MCMC) method. They have analyzed the validity of the generalized laws of thermodynamics on the horizons assuming the Hawking temperature of the horizons. However, in this paper we have used the KHT for our analysis. This approach is different from the method used by them. Thus the result of our analysis on the horizons, is more appropriate than those obtained by them. While inspecting the rate of variation of the entropy on the horizon as a function of redshift, they have extended their plot into the future epoch (i.e. to negative values of $z$). In this paper we have also extended our plots into the future epoch.

\subsection{MCG}

The case $ n=0 $ in the VMCG model represents the MCG dominated FRW universe. In FIG.\ref{fig2} we observe that the curve for $ n=0 $ always obeys the GSLT on the cosmological apparent horizon.

\subsection{GCCG}

\begin{figure}[ht]
    \centering
    \subfloat[Subfigure 1 list of figures text][]
        {
        \includegraphics[width=0.45\textwidth]{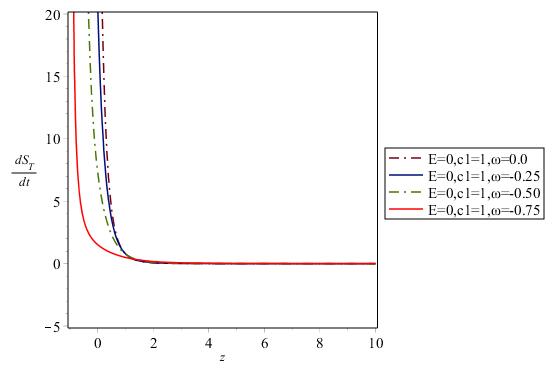}
        \label{fig:subfig3}
        }
    \subfloat[Subfigure 2 list of figures text][]
        {
        \includegraphics[width=0.48\textwidth]{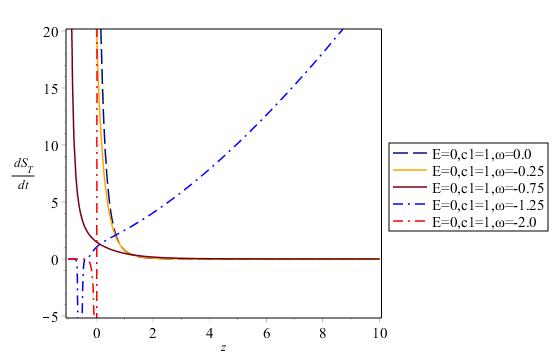}
        \label{fig:subfig4}
        }
    \caption{Variation of $ \dfrac{dS_{T}}{dt} $ wrt the free parameters $ \omega$ maintaining the validity of GSLT on the cosmological apparent horizon for GCCG.}
    \label{fig3}
\end{figure}

In \cite{SS}, Sharif et al analyzed the thermodynamic stability of GCCG. Here we examine the validity of GSLT on the apparent horizon of the universe filled with GCCG. FIG.\ref{fig3} shows the variation of $ \dfrac{dS_{T}}{dt} $ with respect to the parameter $w$ in the EOS given in (\ref{gccg}), for the FRW universe filled with GCCG. We find that the total entropy always increases when $ \omega>-1 $, but for $ \omega<-1 $, the GSLT is violated in the future epoch.



\begin{figure}[ht]
    \centering
    \subfloat[Subfigure 1 list of figures text][]
        {
        \includegraphics[width=0.45\textwidth]{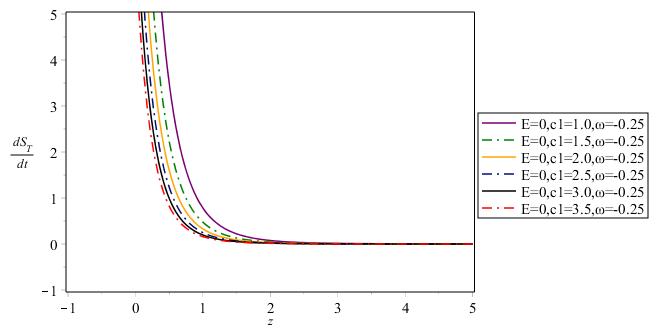}
        \label{fig:subfig5}
        }
    \subfloat[Subfigure 2 list of figures text][]
        {
        \includegraphics[width=0.47\textwidth]{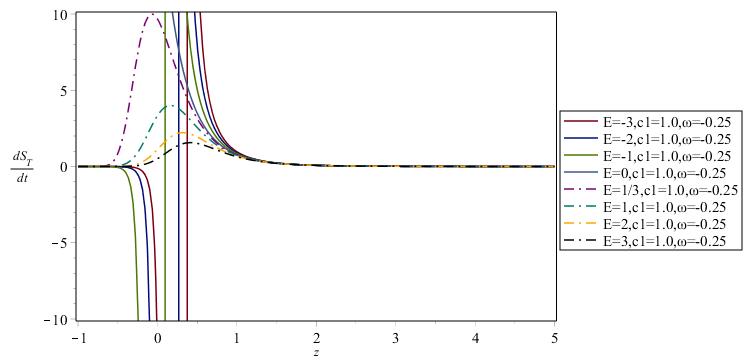}
        \label{fig:subfig6}
        }
    \caption{Variation of $ \dfrac{dS_{T}}{dt} $ wrt the free parameters $ \omega$ maintaining the validity of GSLT on the cosmological apparent horizon for GCCG.}
    \label{fig4}
\end{figure}

Next we study the variation of $ \dfrac{dS_{T}}{dt} $ with respect to the parameter $w$ of the GCCG dominated FRW universe for different values of the integration constant $ c_{1} $, and examine its variation in FIG.\ref{fig4}(a). From this figure it is clear that the GSLT is valid on the cosmological apparent horizon only for positive $ c_{1} $.

We also study the variation of $ \dfrac{dS_{T}}{dt} $ with respect to the parameter $w$ for a variation in $E$. This is shown in FIG.\ref{fig4}(b). It is evident that the GSLT is valid on the apparent horizon only for $ E>0 $. So we conclude that for the validity of GSLT on the cosmological apparent horizon, the GCCG model parameters must satisfy the conditions $ w>-1, E\geq 0$, and $ c_{1}>0 $.

\subsection{MCCG}
\begin{figure}[ht]
    \centering
    \subfloat[Subfigure 1 list of figures text][]
        {
        \includegraphics[width=0.50\textwidth]{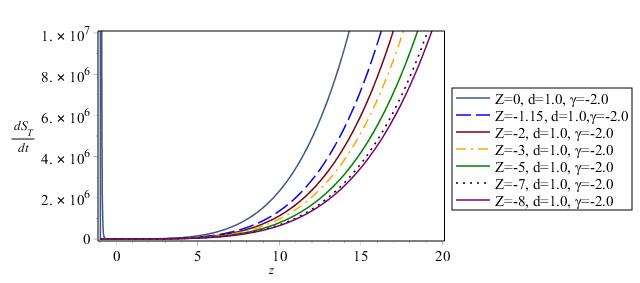}
        \label{fig:subfig7}
        }
    \subfloat[Subfigure 2 list of figures text][]
        {
        \includegraphics[width=0.45\textwidth]{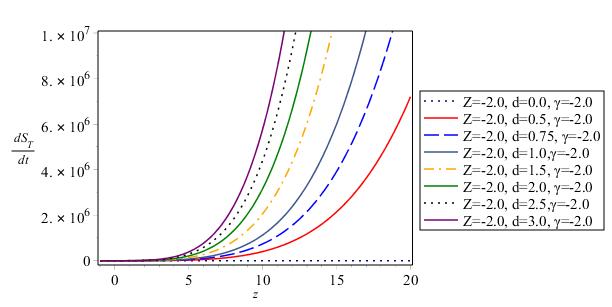}
        \label{fig:subfig8}
        }
    \caption{Variation of $ \dfrac{dS_{T}}{dt} $ wrt the free parameters $ Z,\gamma $ and the integration constant $d$, while preserving the validity of GSLT on the cosmological apparent horizon for MCCG, where we have chosen $ \alpha=0.1, \: \textrm{and} \: A=2 .$}
    \label{fig5}
\end{figure}

We now study the variation of $ \dfrac{dS_{T}}{dt} $ with respect to the arbitrary parameter $Z$ in the EOS of the MCCG dominated FRW universe and show the corresponding variation in the subsequent figures. In the FIG.\ref{fig5}(a), we find that the entropy of FRW universe filled with MCCG is well behaved and obeys the GSLT for negative $ Z $, and from FIG.\ref{fig5}(b) we find that the GSLT is valid for positive integration constant $ d $.

\begin{figure}[ht]
    \centering
    \subfloat[Subfigure 1 list of figures text][]
        {
        \includegraphics[width=0.46\textwidth]{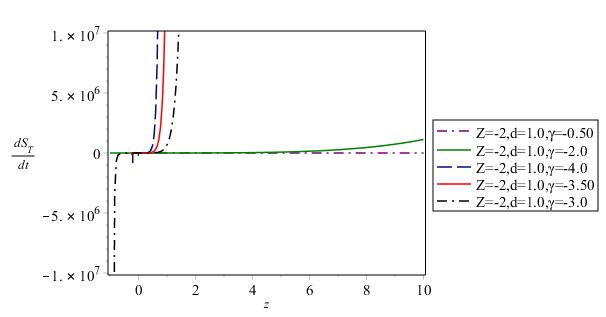}
        \label{fig:subfig9}
        }
    \subfloat[Subfigure 2 list of figures text][]
        {
        \includegraphics[width=0.48\textwidth]{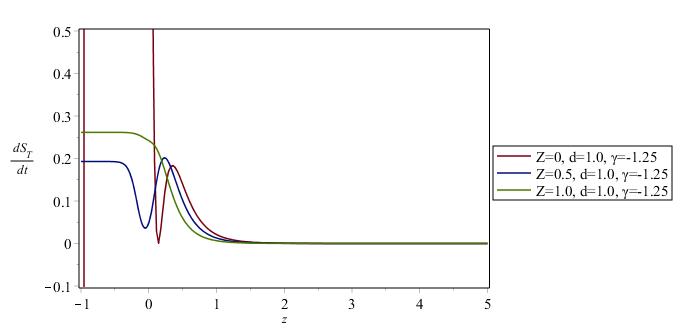}
        \label{fig:subfig10}
        }
    \caption{Variation of $ \dfrac{dS_{T}}{dt} $ wrt the free parameters $ Z,\gamma $ and the integration constant $d$, while preserving the validity of GSLT on the cosmological apparent horizon for MCCG.}
    \label{fig6}
\end{figure}

FIG.\ref{fig6}{\bf{(a)}} depicts some more variations of the parameters and evidently the total entropy is well behaved only around $ Z=-2,\gamma=-2 $ and $ d=1 $. Here we have assumed the values of the parameters as  $\alpha=0.1$, and $ A=2$. In FIG.\ref{fig6}(b), we have explored the variations for different values of fixed parameters i.e. $ A=1$, and $ \alpha=1 $. We can see that the GSLT is valid in this case except for $ Z=0, d=1.0, \: \textrm{and} \: \gamma=-1.25 $. We also find that the GSLT is valid for positive $ Z $. Thus we conclude that the parameter $ d $ must be positive and accordingly there are distinct values of $ \gamma $ for the validity of GSLT on the cosmological apparent horizon in the MCCG model.

\subsection{NVMCG}

\begin{figure}[ht]
\includegraphics[width=0.63\textwidth]{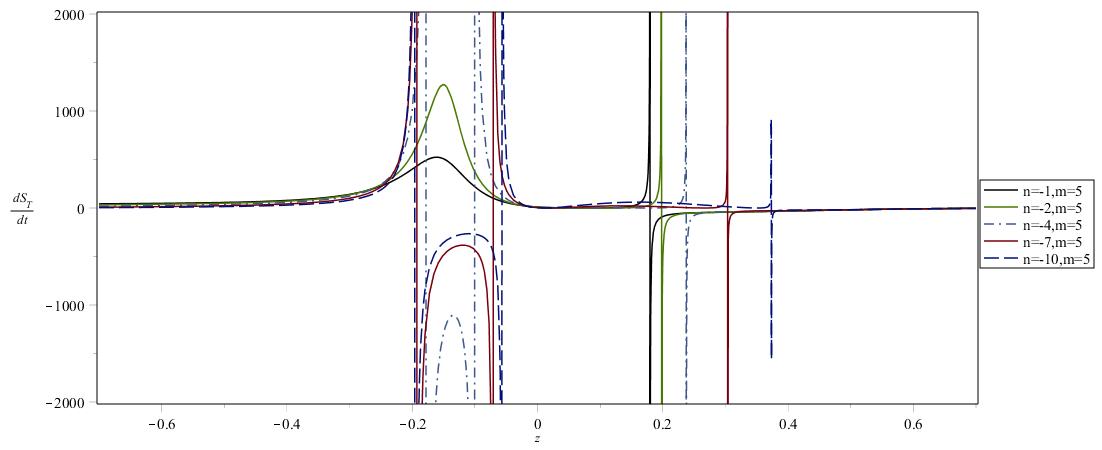}
\caption{Variation of $ \dfrac{dS_{T}}{dt} $ as a function of redshift on the cosmological apparent horizon for different values of  $ n $ with $ m=5 $ for NVMCG.}
\label{nvmcgnvary1}
\end{figure}

\begin{figure}[ht]
\includegraphics[width=0.63\textwidth]{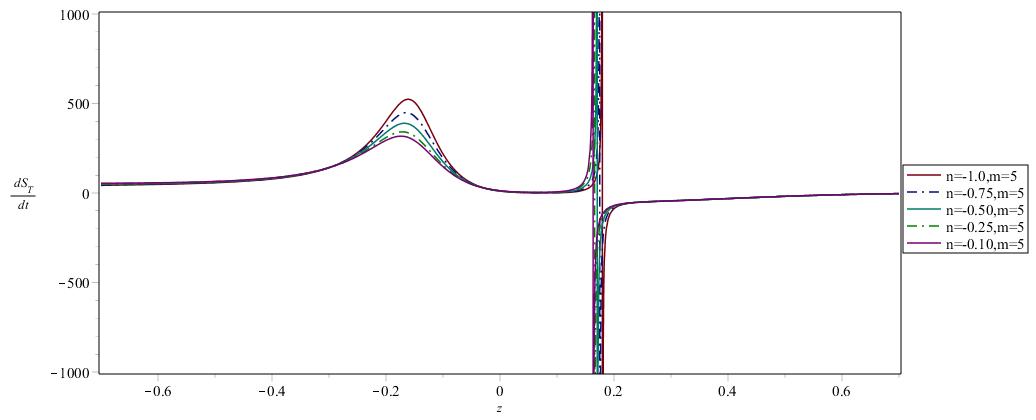}
\caption{Variation of $ \dfrac{dS_{T}}{dt} $ as a function of redshift on the cosmological apparent horizon for different values of  $ n $ with $ m=5 $ for NVMCG.}
\label{nvmcgnvary2}
\end{figure}

\begin{figure}[ht]
\includegraphics[width=0.60\textwidth]{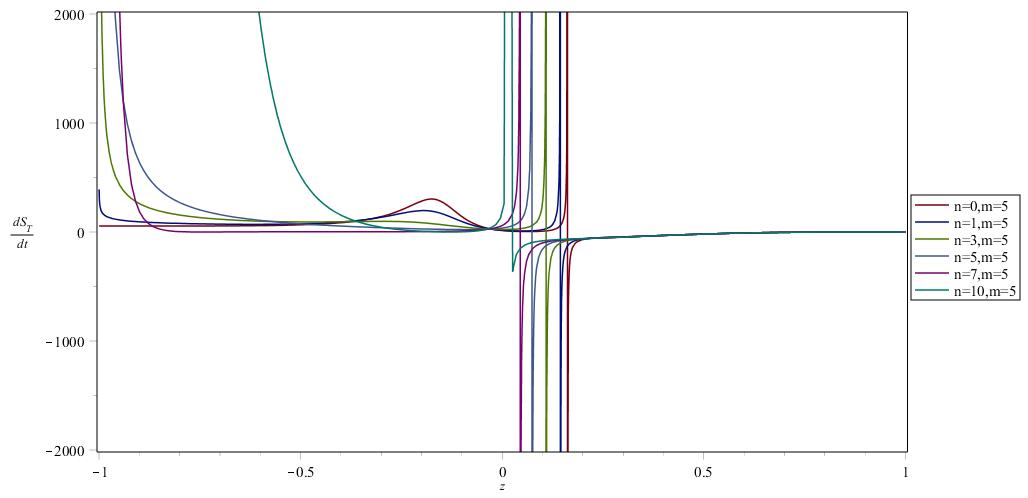}
\caption{Variation of $ \dfrac{dS_{T}}{dt} $ as a function of redshift on the cosmological apparent horizon for different values of  $ n $ with $ m=5 $ for NVMCG.}
\label{nvmcgnvary3}
\end{figure}

In this section we will do a similar analysis for the NVMCG model. In our calculations we have assumed the values $ A_{0}=1.0 $, $ B_{0}=10.0 $, $ \alpha=1.0 $, and $ c_{0}=1.0 $ for the parameters in the EOS of NVMCG. In FIG.\ref{nvmcgnvary1}, FIG.\ref{nvmcgnvary2} and  FIG.\ref{nvmcgnvary3}, we have fixed $ m=5 $ and varied $ n $. We can see that none of the plots are well-behaved and every curve indicates a conditional validity of the GSLT. Moreover all these plots show an abrupt variation in the rate of change of the total entropy during the evolution of the universe, which is difficult to explain.

\begin{figure}[ht]
\includegraphics[width=0.60\textwidth]{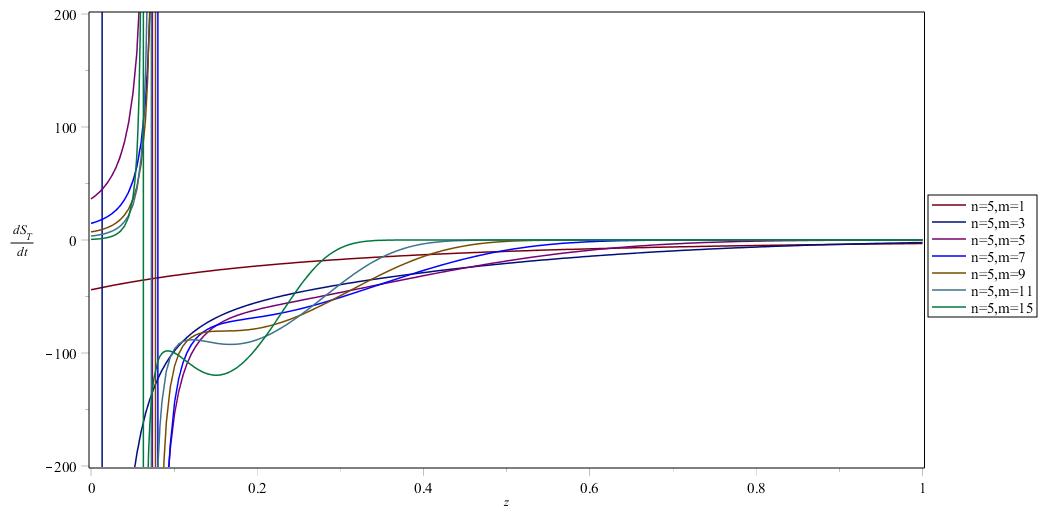}
\caption{Variation of $ \dfrac{dS_{T}}{dt} $ as a function of redshift on the cosmological apparent horizon for different values of $ m $ with $ n=5 $ for NVMCG.}
\label{nvmcgmvary}
\end{figure}

In FIG.\ref{nvmcgmvary} we have fixed the parameter $n$  and varied $ m $. Here also none of the curves satisfy the GSLT over their entire evolution. All the curves violate GSLT in the early phase of the universe and then suddenly the entropy increases in the recent epoch of the universe with vertical asymptotic discontinuities.

Thus from the above analysis we conclude that the NVMCG violates GSLT on the cosmological apparent horizon.

\section{Validity of GSLT on the Cosmological Event Horizon of various Chaplygin gas models}
We now proceed to examine the status of the GSLT on the cosmological event horizon (EH) of FRW universe dominated by the various Chaplygin gas fluids. We like to point out that our analysis on the event horizon is a general treatment without assuming any specific form of temperature.

As the event horizon is `teleological' in nature \cite{JN,HE}, we are only interested in analyzing the overall validity of the GSLT on it. In the calculations with regard to the cosmological EH, we have ignored the computation of $R_{EH}$ and the corresponding graphs, as the calculations become complicated due to increasing number of parameters involved in the EOS of these models. Though the radius of the cosmological event horizon can be calculated using equation (\ref{R_EH}), we do not require its explicit expression to understand the nature of validity of the GSLT on the cosmological EH.  Rather we have tried to examine the validity of GSLT using simple algebraic manipulations which clearly demonstrates the conditional nature of the validity.
However, we have examined the case of the VMCG dominated FRW universe in details as this model is cosmologically significant. Hence we have checked the validity of the GSLT on the cosmological event horizon in the VMCG case.

\subsection{VMCG}
In the case of the Variable Modified Chaplygin gas, the rate of change of total entropy on the cosmological event horizon is given by
\begin{equation}
\frac{dS_{T}}{dt}=\frac{4\pi R_{EH}^{2}H}{T_{EH}}\left(A \rho - \frac{B}{\rho^{\alpha}}+\rho \right)(R_{EH}-R_{AH}).
\end{equation}
Now for the validity of GSLT ($i.e. \dot{S}_{T}>0$), we need $ R_{EH}>R_{AH} $, which implies that we must have $ (A+1)\rho>\frac{B}{\rho^{\alpha}} $. This finally leads us to the condition
\begin{equation}\label{vmcg_cond}
\rho^{1+\alpha} >\frac{B_{0}a^{-n}}{(1+A)}.
\end{equation}

In the above condition (\ref{vmcg_cond}), we substitute the expression for energy density of the VMCG given by (\ref{09}), which is
\begin{equation}
\rho=\frac{\rho_{0}}{a^{\frac{n}{1+\alpha}}}\left[\Omega_{x}+(1-\Omega_{x})\left(\frac{1}{a}\right)^{3N}\right]^{\frac{1}{1+\alpha}}.
\end{equation}
This leads us to the following relation:
\begin{equation}
 \left[\Omega_{x} +(1-\Omega_{x})\frac{1}{a^{3N}}\right] > \frac{N\Omega_{x}}{(1+\alpha)(1+A)}.
\end{equation}

Replacing the scale factor $ a $ by the redshift $ z $ by using the substitution $ a=\frac{1}{z+1} $, we obtain
\begin{equation}\label{gslt_vmcg}
(z+1)^{3N}>\frac{-n\Omega_{x}}{3(1+A)(1+\alpha)(1-\Omega_{x})}.
\end{equation}

Considering the present day case we put $ z=0 $ in the above relation, and arrive at the limiting condition
\begin{equation}\label{limitvmcg}
n>3(1+A)(1+\alpha)\left(1-\frac{1}{\Omega_{x}}\right).
\end{equation}

\begin{figure}[ht]
    \centering
    \subfloat[Subfigure 1 list of figures text][]
        {
        \includegraphics[width=0.33\textwidth]{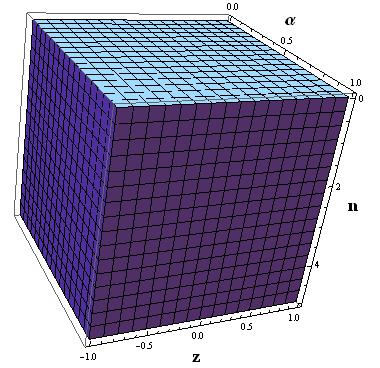}
        \label{fig:subfig11}
        }
     \hspace{0.75in}
    \subfloat[Subfigure 2 list of figures text][]
        {
        \includegraphics[width=0.30\textwidth]{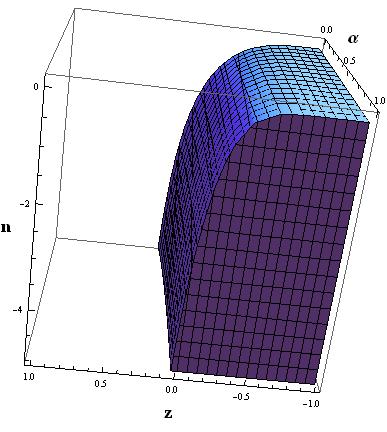}
        \label{fig:subfig12}
        }
    \caption{Variation of $ z $ wrt the free parameters $ n, \alpha $ while preserving the validity of GSLT on the cosmological event horizon for the condition of quintessence (fig. a) and phantom dominated universe (fig. b).}
    \label{fig_parameter}
\end{figure}

The above criterion depicts the condition for the validity of the GSLT on the event horizon of a VMCG dominated FRW universe. Let us assume $ A=1/3 $ , $ \Omega_{x}=0.7 $  and $ \alpha=0.25 $ in order to model a cosmologically viable evolution of the universe. This assumption leads us to the condition
\begin{equation}
(z+1)^{(5-n)}>-\left(\frac{n}{2.14}\right).
\end{equation}
From this relation we can immediately observe the explicit dependence of the redshift $z$ on the free parameter $ n $. We also see that when $ n $ is zero or positive, the relation becomes a trivial one, but when $ n $ becomes negative (signifying a phantom dominated universe), it prevents the redshift $ z $ from attaining the value of $ -1$ (indicating the future of the universe), as the left hand side becomes zero but the right hand side is still positive. This  indicates that for non-negative values of $ n $, $ (p+\rho)>0 $, and therefore, the above relation is only valid for $ n\geq 0 $.
Now putting $z=0  $ in the criterion (\ref{limitvmcg}) and setting  $ A=1/3 $, $ \alpha=0.25 $ and $ \Omega_{x}=0.7 $, we get
\begin{align}
n>3(1+A)(1+\alpha)\left(1-\frac{1}{\Omega_{x}}\right) \qquad \Rightarrow \qquad n>-2.14. \nonumber
\end{align}
Thus we arrive at the condition that $ n>-2.14 $ for the chosen values of the parameters for the validity of GSLT in the VMCG filled FRW universe bounded by the event horizon, provided we assume the \emph{a priori} condition that $ R_{EH}>R_{AH} $. From the cosmological analysis of VMCG in FRW universe by earlier workers, it has been found that for $ n\geq 0 $, the universe is dominated by quintessence \cite{VMCG1} or cosmological constant. Therefore the criterion (\ref{limitvmcg}) obtained above is consistent with it. From the above analysis we can conclude that for $n\geq 0$, GSLT is valid on the event horizon in a VMCG dominated FRW universe (Fig.~\ref{fig_parameter}(a)). However, we note here that the model VMCG itself is thermodynamically unstable in this range of $ n\geq 0 $ \cite{vmcg}.



Another condition for the validity of GSLT is $R_{EH}<R_{AH}$ with $ (p+\rho)<0 $. From the second condition quoted beside, we can say that
\begin{equation}\label{vmcg_another}
(z+1)^{3N}<\frac{-n\Omega_{x}}{3(1+A)(1+\alpha)(1-\Omega_{x})}.
\end{equation}
Now for the appropriate values of the free parameters $A=1/3, \Omega_{x}=0.7$ and $\alpha=0.25 $, equation (\ref{vmcg_another}) leads us to the relation
\begin{equation}\label{vmcg_particular}
(z+1)^{5-n} < - 0.47 n.
\end{equation}
This relation is not valid for $ n\geq0 $. It is known that the condition $ (p+\rho)<0 $ represents the case for $ n<0 $ (phantom dominated universe), which is consistent with the above relation (\ref{vmcg_particular}). It is easy to realize that for $ -1\leq z \leq 0 ,$ the relation certainly holds true, and for $ z=0 $ it yields us the condition $ n>-2.14 $. In fact, this condition is true up to some positive values of $ z $ depending on the corresponding value of $ n $. So we can certainly say that the GSLT holds for the FRW universe filled with the VMCG fluid during the current epoch and will also hold in the future.

However we can also see that for positive high values of $ z $ (in the early universe), the relation (\ref{vmcg_particular}) fails to hold. Consequently the GSLT is not valid in the early universe for a Phantom like VMCG (Fig.~\ref{fig_parameter}(b)).

Therefore we can safely claim that for $ n<0 $ (i.e. $R_{EH}<R_{AH}  $), the VMCG dominated FRW universe violates the GSLT on the event horizon in the early phase of the universe but it holds during the current epoch and will also hold in the future, and moreover in this range of $ n<0 $, the VMCG model itself is thermodynamically stable \cite{vmcg}.

At this juncture we like to note that in our other paper \cite{CGP}, we have checked the temperature variation of the FRW universe filled with VMCG, which is consistent with the thermodynamic dependence of the values of the parameter $n$. In the present paper also we find that the validity of the GSLT on the apparent horizon depends very much on the value of the parameter $n$, when we use the temperature of VMCG dominated FRW universe in place of the horizon temperature, for conditions approaching thermal equilibrium. Our analysis clearly shows that negative value of $n$ is thermodynamically stable (or favoured) in the sense that the GSLT is valid on the apparent horizon in such cases. In the same line of thinking we tried to investigate the validity of the GSLT on the event horizon of the VMCG universe and found that the parameter space of the GSLT validity condition for the phantom case is incomplete, but is complete for the quintessence case (Fig.~\ref{fig_parameter}).

\subsection{MCG}

Substituting $ n=0 $ in the equation of state for VMCG, we get the equation of state for MCG:

\begin{equation}
p=A\rho -\frac{B}{\rho^{\alpha}}.
\end{equation}
The condition of applicability of GSLT on the event horizon of a MCG dominated FRW universe is then
\begin{equation}\label{gslt_mcg}
(z+1)^{3(1+A)(1+\alpha)}>0
\end{equation}
which is a trivial relation given the equation of state of MCG. Now if we consider  $z=0$ for our present day, then by putting $ n=0 $ in the relation for $n $, we get
\begin{equation}
3(1+A)(1+\alpha)\left(1-\frac{1}{\Omega_{x}}\right)<0.
\end{equation}
For the chosen values of the parameters $A$, $\alpha$ and $\Omega_x$ mentioned in the earlier section, we find that the first two terms in the equation of state of MCG are positive, and we know that the third term must be negative. Therefore we can say that the GSLT is always valid on the event horizon of a MCG dominated universe iff $ R_{EH}>R_{AH} $.

\subsection{GCG}
The equation of state for GCG is given in (\ref{gcg_eos}) as
\begin{equation}
p=-\frac{B}{\rho^{\alpha}}.
\end{equation}
To consider the case of Generalized Chaplygin gas, we substitute $ n=0 $ and $ A=0 $ in the viability relation (\ref{gslt_vmcg}) and we obtain
\begin{equation}
 (z+1)^{3(1+\alpha)}>0,
\end{equation}
which is again a trivial condition. Moreover, if we put $ n=0 $ and $ A=0 $ in the relation (\ref{limitvmcg}) for $ n $, then for the present day value of $ z=0 $, we get
\begin{equation}\label{limitgcg}
 3(1+\alpha)\left(1-\frac{1}{\Omega_{x}}\right)<0.
\end{equation}
We know that the first term is always positive in the equation of state of GCG and the second term is always negative. Therefore the above condition (\ref{limitgcg}) is also trivial. Therefore we can say that the GSLT is always valid on the event horizon of a GCG dominated universe iff $ R_{EH}>R_{AH} $.

\subsection{GCCG}

The equation of state for the GCCG quoted in Section II is given by \cite{GCCG}
\begin{equation}
p=-\rho^{-\alpha}[c+(\rho^{\alpha+1}-c)^{-w}],
\end{equation}
and the expression for the energy density is
\begin{equation}
\rho=[c+(c_{1}NV^{-N}+1)^{\frac{1}{w+1}}]^{\frac{1}{\alpha+1}},
\end{equation}
where $c_{1}  $ is an arbitrary integration constant, and $ N=(1+\alpha)(1+w) $. The rate of change of total entropy in this model obeys GSLT on the cosmological event horizon if $ R_{EH}>R_{AH} $, and $ (p+\rho)>0 $. From the second condition beside, we get
\begin{align}
 \frac{c_{1}NV^{-N}}{(c_{1}NV^{-N}+1)^{\frac{w}{w+1}}}>0.
\end{align}
From the above expression we can say that the GSLT is valid in this case if $ c_{1}N>0 $. So there are two possibilities: either (i) $c_{1}>0  $ and $N>0  $, or (ii) $c_{1}<0  $ and $N<0  $.

But we know that $ (1+\alpha)>0 $. Therefore the condition $ N>0 $ means that $ -1<w<0 $, and when $ N<0 $, it means that $ w<-1 $. However the case $ N<0 $ is not possible because the equation of state parameter in this case should be greater than $ -1 $. Therefore when $ R_{EH}>R_{AH} $, the condition for the GSLT to be valid on the event horizon of a FRW universe filled with GCCG is $c_{1}>0  $, and $ -1<w<0 $. The initial conditions have to be chosen in such a manner that we have $c_{1}>0 $.

The other condition for the validity of the GSLT on the event horizon of this universe is $ R_{EH}<R_{AH} $ and $ (p+\rho)<0 $, which will then represent the phantom case. In the same way as before, we arrive at the relation
\begin{equation}
\frac{c_{1}NV^{-N}}{(c_{1}NV^{-N}+1)^{\frac{w}{w+1}}}<0,
\end{equation}
which is only possible when $c_{1}N<0  $. Now we know that for the phantom case, the equation of state parameter is $ w<-1 $, which suggests that $ N<0 $. So $ c_{1} $ must be positive.

Therefore when $ R_{EH}>R_{AH} $, the GSLT is valid when $c_{1}>0  $ and $ -1<w<0 $, and when $ R_{EH}<R_{AH} $,
the conditions are $c_{1}>0  $ and $ w<-1 $. So, if we choose the boundary conditions in such a way that $c_{1}>0  $, then the GCCG dominated FRW universe obeys GSLT on the event horizon.

\subsection{MCCG}

The equation of state for the Modified Cosmic Chaplygin gas is given by
\begin{equation}
P=A\rho-\rho^{-\alpha}[(\rho^{\alpha+1}-C)^{-\gamma} + C],
\end{equation}
where $ 0<\alpha\leq 1  $, $ -b<\gamma<0 $, $ b\neq 1  $, $ C=\frac{Z}{\gamma +1} -1 $, $ Z $ being an arbitrary constant, and $ A $ is a positive constant. The approximate form of energy density is
\begin{equation}
\rho=\left[ \dfrac{C+(-C)^{-\gamma}+(\frac{\varepsilon}{V})^{M}}{A+1+\gamma(-C)^{-\gamma-1}}\right]^{\frac{1}{1+\alpha}},
\end{equation}
where $ \varepsilon=d(A+1)^{\frac{1}{M}} $, $ M=(1+\alpha)(1+A) $, $ A+1+\gamma(-C)^{-\gamma-1}\neq 0 $, and $ d $ is the constant of integration. Now from the EOS of MCCG, we conclude that $ (\rho^{\alpha +1}-C)>0 $, which then leads us to the condition
\begin{equation}
\rho^{\alpha +1}>\dfrac{Z-(\gamma +1)}{(\gamma +1)}.
\end{equation}
For $ C>0 $, we have $ Z>(\gamma+1) $. In this case, if $ -1<\gamma<0 $, i.e. $ Z>0 $, then $ \rho^{\alpha +1} $ must be greater in magnitude than some positive constant, but if $ \gamma<-1 $, no specific conclusion can be drawn regarding the value of $ \rho $. Again when $ C<0 $, which means that $ Z<(\gamma +1) $, then $ \rho^{\alpha +1} $ is greater than some positive constant if $ \gamma<-1 $ (i.e. $ Z<0 $), but is greater than a negative value if $ -1<\gamma<0 $ (i.e. $ Z<1 $).

From the expression of energy density we can say that $ \gamma $ can only take integral values, and so we can discard the $ -1<\gamma<0 $ limit.
Clearly we can see that if $ C<0 $ and $ \gamma<-1 $ (i.e. $ Z<0 $), we can always have a lower positive bound of $ \rho $.

To check for the validity of the GSLT on the event horizon of MCCG dominated FRW universe, we note that the criterion for this validity is $ (P+\rho)>0 $ when $ R_{EH}>R_{AH} $, as in the previous cases. Using this inequality we obtain the following condition:
\begin{equation}
(A+1)\rho^{\alpha +1}-[(\rho^{\alpha+1}-C)^{-\gamma}+C]>0.
\end{equation}
Substituting the expression for energy density and using binomial expansion, we get
\begin{align}
 \left(\frac{\varepsilon}{V}\right)^{M}+(-C)^{-\gamma}-\rho^{\alpha+1}\gamma(-C)^{-\gamma-1}
  -[(-C)^{-\gamma}+(-\gamma)\rho^{\alpha+1}(-C)^{-\gamma-1}+...]>0,
\end{align}
and the end result is
\begin{equation}
\left(\frac{\varepsilon}{V}\right)^{M}>[\rho^{(\alpha+1)(-\gamma)}+(-\gamma)\rho^{(\alpha+1)(-\gamma-1)}(-C)+...].
\end{equation}
The right hand side is definitely a positive quantity if $ C<0 $. Denoting the entire term inside the square brackets by $ K $, we find that $ K>0 $ for $ C<0  $, but for $ C>0 $ nothing can be said regarding the sign of $K$. Also previous analyses of the equation of state of MCCG in cosmology showed that for negative $ C $ and $ Z<0 $, there is a positive lower bound for the energy density, and hence for the condition of validity of GSLT on the event horizon of MCCG we must have
\begin{equation}
d>V\left[ \frac{K}{(A+1)} \right]^{\frac{1}{M}}.
\end{equation}
This gives us a positive lower bound for the integration constant $d$. Therefore we can say that for $ C<0 $ and $ \gamma<-1 $ (or $ Z<0 $), the GSLT is valid on the event horizon of a MCCG dominated FRW universe when we choose our initial conditions in such a way that the integration constant stays above the lower bound $V\left[ \frac{K}{(A+1)} \right]^{\frac{1}{M}}$. In this limit of $ Z <0$, the model MCCG itself is also thermodynamically stable \cite{SA}.

\subsection{NVMCG}
In this section we analyze the the validity of GSLT for the NVMCG model. The equation of state for the New Variable Modified Chaplygin gas is already quoted earlier and is given by
\begin{equation}
p=A(a)\rho - \frac{B(a)}{\rho^{\alpha}},
\end{equation}
and the expression for energy density is
\begin{align}
\rho&=a^{-3}e^{\frac{3A_{0}a^{-m}}{m}} \bigg[c_{0}+\frac{B_{0}}{A_{0}}\left( \frac{3A_{0}(1+\alpha)}{m} \right)^{\frac{3(1+\alpha)+m-n}{m}} \times \Gamma \left(\frac{n-3(1+\alpha)}{m},\frac{3A_{0}(1+\alpha)a^{-m}}{m}\right)\bigg]^{\frac{1}{1+\alpha}},
\end{align}
where $ \Gamma(x,y) $ is the upper incomplete gamma function and $ c_{0} $ is the integration constant.
The rate of change of total entropy obeys GSLT on the cosmological event horizon if $ R_{EH}>R_{AH} $, and $(p+\rho)>0$. From the second condition we get the following relation
\begin{equation}
\rho^{\alpha +1}>\frac{B_{0}a^{-n}}{A_{0}a^{-m}+1}.
\end{equation}
It follows that if $ n>0 $ and $ m>0 $ in the limit of large $ a $, the above relation reduces to the form $\rho^{\alpha +1}>0  $,
which can correspond to the `quintessence' form of dark energy, whereas for $ n<0 $ and $ m>0 $, the relation becomes $\rho^{\alpha +1}>\infty  $, so that the energy density blows up, which corresponds to the phantom model of dark energy.

Now for the GSLT to be valid, the necessary condition is $ (p+\rho)>0 $, which means that the equation of state parameter is $ w>-1 $ (which corresponds to the quintessence model). Therefore we can say that in the NVMCG model, the GSLT is valid on the cosmological event horizon if $ n>0 $ and $ m>0 $ along with $ R_{EH}>R_{AH} $.

The other condition for the validity of GSLT on the cosmological event horizon is as usual $ (p+\rho)<0 $, when $ R_{EH}<R_{AH} $. The condition $ (p+\rho)<0 $  can be written as
\begin{equation}
\rho^{\alpha+1}<\frac{B_{0}a^{-n}}{A_{0}a^{-m}+1}.
\end{equation}
If we consider the case $ n>0 $ and $ m>0 $ in the large $ a $ limit, the above relation becomes $\rho^{\alpha+1}<0 $, which is not physically possible. So we can safely discard this case.

Next we consider the range $ n<0 $ and $ m>0 $, which yields the relation $\rho^{\alpha+1}<\infty  $, which is a perfectly acceptable criterion. Using the expression for energy density we arrive at the condition
\begin{align}
c_{0} > \left[ \frac{B_{0}a^{-(n-3(1+\alpha))}}{(A_{0}a^{-m}+1)} \right]e^{\frac{-3A_{0}(1+\alpha)a^{-m}}{m}}
  -\frac{B_{0}}{A_{0}}\left(\frac{3A_{0}(1+\alpha)}{m}\right)^{\frac{m-(n-3(1+\alpha))}{m}} \Gamma\left(\frac{n-3(1+\alpha)}{m},\frac{3A_{0}(1+\alpha)a^{-m}}{m}\right).
\end{align}
This relation implies that $ m $ cannot be negative. When $ m\sim 0 $, the NVMCG model asymptotically approaches the VMCG model and for small positive values of $ m $, the above condition becomes
\begin{align}
c_{0} & > \left[ \frac{B_{0}a^{-(n-3(1+\alpha))}}{(A_{0}a^{-m}+1)} \right]e^{\frac{-3A_{0}(1+\alpha)a^{-m}}{m}} \nonumber \\
 & - \frac{B_{0}}{A_{0}}\left(\frac{3A_{0}(1+\alpha)}{m}\right)^{\frac{m-(n-3(1+\alpha))}{m}} \left[\left(\frac{3A_{0}(1+\alpha)a^{-m}}{m}\right)^{\frac{n-3(1+\alpha)}{m}-1} e^{-\frac{3A_{0}(1+\alpha)a^{-m}}{m}}\right],
\end{align}
that is
\begin{align}
c_{0}>\left[\frac{B(a)}{(A(a)+1)}-\frac{B(a)}{A(a)}\right]a^{3(1+\alpha)} e^{-\frac{3A_{0}(1+\alpha)a^{-m}}{m}}. \nonumber
\end{align}
For large $a$ and small positive $ m $ approaching zero, we find that the limit for $ c_{0} $ in the NVMCG model is
\begin{equation}
c_{0}>\left[\frac{B_{0}}{(A_{0}+a^{m})}-\frac{B_{0}}{A_{0}}\right]a^{3(1+\alpha)-n+m}.
\end{equation}
Thus the limit on $ c_{0} $ depends explicitly on the exponent $3(1+\alpha)-n+m  $. If $ n<3(1+\alpha)+m $, the value of the integration constant blows up (i.e. $ c_{0}>-\infty $) implying that the energy density blows up for large values of $ a $. But if $ n>3(1+\alpha)+m $, we must have $ c_{0}>0 $ in the large $ a $ limit. In the special case $ n=3(1+\alpha)+m $, the integration constant becomes $ c_{0}>\left[\frac{B_{0}}{(A_{0}+a^{m+3(1+\alpha)})}-\frac{B_{0}}{A_{0}}\right] $, i.e. it is greater than a negative constant value.
In the asymptotic limit of $ m\sim 0 $, we obtain the condition $ c_{0}>0 $ for large values of $ a $, for the validity of GSLT in the FRW universe filled with NVMCG and bounded by the event horizon.

\section{Discussions}

The entire analysis in this paper is based on the assumption that there is enough time for the fluid to attain thermal equilibrium with the cosmological horizons. If there is no thermal equilibrium, the validity of GSLT on the cosmological horizons becomes much more conditional.

In absence of thermal equilibrium between the fluid and the horizon \cite{TB} (i.e. $ T_{H}\neq T_{b} $), we may assume a near equilibrium situation and write an approximate relation for the rate of change of total entropy as
\begin{equation}
\frac{dS_{T}}{dt}=4\pi R_{H}^{2}(p+\rho)\left( \frac{H R_{H}}{T_{H}} + \frac{(\dot{R_{H}} - H R_{H})}{T_{b}} \right).
\end{equation}
We can always think of the bulk temperature to be very near to the horizon temperature i.e. $T_{b}=T_{H}+\delta T$, where we are assuming $ \dfrac{\delta T}{T_{H}}<<1 $. The rate of change of total entropy in near-equilibrium situation becomes
\begin{equation}
\dfrac{dS_{T}}{dt}\simeq \dfrac{dS^{0}_{T}}{dt} + \dfrac{\delta T}{T^{2}_{H}} 4\pi R^{2}_{H}(p+\rho)(HR_{H}-\dot{R_{H}}),
\end{equation}
where $ \dfrac{dS^{0}_{T}}{dt}\equiv \frac{4\pi R_{H}^{2}(p+\rho)\dot{R}_{H}}{T_{H}} $, is the rate change of total entropy when there is thermal equilibrium.

For the cosmological apparent horizon (with $ T_{AH}\neq T_{b} $), the rate of change of total entropy becomes
\begin{equation}
\frac{dS_{T}}{dt}= \frac{4\pi (p+\rho)^{2}}{2H^{4}T_{b}} + \left(\frac{4\pi}{H^{2}T_{AH}T_{b}} \right) (p+\rho)(T_{b}-T_{AH}).
\end{equation}
In that case, we can see that the GSLT is not always valid on the cosmological apparent horizon. The first term on the right hand side is always positive, and therefore either the second term is also positive (which suggests that $ (p+\rho)>0$ for $T_{b}>T_{AH}$, or $(p+\rho)<0,T_{b}<T_{AH}  $), or as a whole the right hand side of the above equation remains positive even if the second term becomes negative.

If $(p+\rho)>0  $, and $ T_{b}<T_{AH} $, we obtain the condition
\begin{equation}
\frac{(p+\rho)}{2H^{2}}>\left(1-\frac{T_{b}}{T_{AH}}\right),
\end{equation}
which represents the quintessence dominated universe. Another possibility is that if $(p+\rho)<0 $, and $ T_{b}>T_{AH} $, the condition for the validity of GSLT on the cosmological apparent horizon becomes
\begin{equation}
\frac{(p+\rho)}{2H^{2}}<\left(1-\frac{T_{b}}{T_{AH}}\right),
\end{equation}
which then represents the phantom dominated universe.

Now in the near-equilibrium scenario, the rate of change of total entropy on the cosmological apparent horizon can be written as
\begin{equation}
\dfrac{dS_{T}}{dt}\simeq\dfrac{dS^{0}_{T}}{dt} + \dfrac{\delta T}{T_{AH}^{2}}\left[\dfrac{4\pi(p+\rho)}{H^2} - \dfrac{4\pi (p+\rho)^{2}}{2H^{4}}\right] -\dfrac{4\pi(p+\rho)}{H^{2}}\dfrac{(\delta T)^{2}}{T^{3}_{AH}},
\end{equation}
which agrees with the general formula in the first order terms of $ \delta T $. Here $ \dfrac{dS^{0}_{T}}{dt} \equiv \frac{4\pi (p+\rho)^2}{2 H^{4}T_{AH}} $, is the rate of change of entropy on the cosmological apparent horizon when there is thermal equilibrium.

In the same way, if the temperature on the cosmological event horizon and the bulk fluid temperature are not in equilibrium, then we have the rate of change of total entropy as
\begin{equation}
\frac{dS_{T}}{dt}=4\pi R_{EH}^{2}H (p+\rho)\left(\frac{R_{EH}}{T_{EH}} - \frac{R_{AH}}{T_{b}}\right).
\end{equation}
In this case, for the GSLT to be valid on the cosmological event horizon, the conditions are either $ (p+\rho)>0,\frac{R_{EH}}{T_{EH}} > \frac{R_{AH}}{T_{b}}   $ (which represents the quintessence dominated universe) or the other possibility is that $ (p+\rho)<0,\frac{R_{EH}}{T_{EH}}< \frac{R_{AH}}{T_{b}}   $ (the phantom dominated universe).

In the near-equilibrium scenario, the rate of change of total entropy on the cosmological event horizon can be written as
\begin{equation}
\dfrac{dS_{T}}{dt}\simeq\dfrac{dS^{0}_{T}}{dt} + \dfrac{\delta T}{T^{2}_{EH}} 4\pi R^{2}_{EH}(p+\rho),
\end{equation}
which again agrees with the general formula given above. Once again, $ \dfrac{dS^{0}_{T}}{dt} \equiv \frac{4\pi R_{EH}^{2}}{T_{EH}}(p+\rho)(HR_{EH}-1)$ is the rate of change of entropy on the cosmological event horizon when there is thermal equilibrium.

Therefore in dynamical situations, if we assume that the difference in temperature is not very large, and do perturbative analysis around the horizon temperature, we can always get the equilibrium term in the leading order with some correction terms with higher orders in $ \delta T $. Hence, the study of the near equilibrium scenario becomes important from this point of view.

Thus we can see that in absence of thermal equilibrium, the validity of the GSLT on the cosmological horizons becomes far more conditional. Here also we can do the same analysis as done in our previous section to find out how the free parameters affect these conditions.

\section{Conclusions}

In this work we have examined the thermodynamic viability of some dark energy models which may be considered as alternatives to the $\Lambda$CDM model. The Chaplygin gas models, especially the VMCG model, is successful in explaining all three phases of evolution of the universe, namely, the radiation dominated phase, the matter dominated phase and the vacuum energy dominated phase. Thus it is more versatile than the $\Lambda$CDM model. Although strict constraints cannot be obtained from our analysis on the validity of GSLT for the different Chaplygin gas models, yet it provides us with clear ranges of parameters and in some cases these ranges conforms to the results obtained by other authors on those models, thereby establishing the appropriateness of our thermodynamic analysis of these models. Our analysis provides us with a clear picture on how to choose these models if the GSLT is to hold in a FRW universe filled with such fluids. It is to be noted that the universal thermodynamics is not the only factor determining the physics of these models. The validity of GSLT on cosmological horizons is a necessary condition for any cosmological model to be considered physically realistic.

In the case of the cosmological apparent horizon, from our consideration of the Kodama-Hayward temperature, we conclude that the MCG and VMCG models always obey the GSLT. For the case of the GCCG, the model parameters should lie in the range $ c_{1}>0, E\geq 0, -1<w<0 $, whereas for the MCCG model the parameter $ d $ must be positive. However, the NVMCG model does not obey the GSLT on the cosmological apparent horizon during the entire evolution of the cosmos. Either it has abrupt discontinuities, or the total entropy simply decreases in magnitude during certain phases of the evolution. Therefore the NVMCG violates the GSLT on the cosmological apparent horizon. 

For the cosmological event horizon, we find that the validity of GSLT is always conditional, which we enlist below:
\begin{itemize}
  \item For the VMCG model we have shown that for $ n>0 $, the GSLT is valid on the cosmic event horizon with the automatic condition $ R_{EH}>R_{AH} $. For $ n<0 $, with the condition $ R_{EH}<R_{AH} $, the GSLT is violated on the cosmological event horizon (i.e. phantom dominated VMCG model violates GSLT on the event horizon). Therefore the validity of the GSLT on the event horizon favors the quintessence dominated ($n>0$) FRW universe in the VMCG model.
  \item In the MCG dominated FRW universe, the GSLT is always valid on the cosmological event horizon. It is also valid for GCG dominated FRW universe.
  \item In the case of GCCG dominated FRW universe, the GSLT is valid conditionally on the cosmological event horizon. In the case when $R_{EH}>R_{AH}  $,  the equation of state parameter has to be $ -1<w<0 $, and the integration constant $ c_{1} $ has to be chosen positive. In the case of $ R_{EH}<R_{AH} $, the parameter should be $ w<-1 $, and again the integration constant $ c_{1} $ has to be positive. So in both the cases, depending on the value of the parameter $ w $ and the initial conditions for $ c_{1} $ to be positive, the validity of GSLT on the cosmological event horizon for the GCCG model can be achieved in a FRW universe.
  \item In the NVMCG model, the GSLT is valid on the cosmological event horizon for two conditions. One is that if $R_{EH}>R_{AH}  $, then the GSLT is valid for the parameters $ n>0 $ and $ m>0 $. Therefore when $R_{EH}>R_{AH}  $, the validity of GSLT on the cosmological event horizon favors the quintessence dominated FRW universe for the NVMCG model. The second possible condition is obtained for $ R_{EH}<R_{AH} $. In this case we have shown that the only possibility is $ n<0 $ and $ m>0$ for the validity of the GSLT.
  \item We also want to point out that the limit on the value of $Z$ for the validity of GSLT on the event horizon of the universe filled with MCCG is consistent with the bound on $Z$ as obtained by Sharif \cite{SA} for the thermodynamic stability of MCCG. Both the analysis of \cite{SA} and that of ours yield the condition that $Z<0$.
\end{itemize}
Therefore as we always know that if we consider the universe to be bounded by the cosmological apparent horizon, then every fluid model satisfies the GSLT, but with the cosmological event horizon as the boundary surface, the situation changes, and different models demand different parameter ranges for the validity of the GSLT on the bounding surface.

\section*{Acknowledgments}
The authors are thankful to the reviewers for their valuable comments and suggestions. SC is grateful to CSIR, Government of India for providing junior research fellowship. SG gratefully acknowledges IUCAA, India for an associateship and CSIR, Government of India for approving the major research project No. 03(1446)/18/EMR-II.

\end{document}